\documentstyle[12pt]{article}
\begin{document}
\vspace*{-1in}
\renewcommand{\thefootnote}{\fnsymbol{footnote}}
\begin{flushright}
TIFR/TH/95-57\\
IISc-CTS-6/95\\
hep-ph/9511433 \\
\end{flushright}
\vskip 65pt
\begin{center}
{\Large \bf \boldmath $J/\psi$ production via fragmentation at
HERA} \\
\vspace{8mm}
{\bf R.M.~Godbole\footnote{rohini@cts.iisc.ernet.in},}\\
\vspace{10pt}
{\it Centre for Theoretical Studies, Indian Institute of Science,
Bangalore-560012, India.}\\
\vspace{20pt}
{\bf D.P.~Roy\footnote{dproy@theory.tifr.res.in}}
and {\bf K. Sridhar\footnote{sridhar@theory.tifr.res.in}}\\
\vspace{10pt}
{\it Theory Group, Tata Institute of Fundamental Research, \\
Homi Bhabha Road, Bombay 400 005, India.}

\vspace{80pt}
{\bf ABSTRACT}
\end{center}
\vspace{12pt}

We compute the contributions to large-$p_T$ $J/\psi$ production
at HERA coming from fragmentation of gluons and charm quarks.
We find that the charm quark fragmentation contribution dominates
over the direct production of $J/\psi$ via photon-gluon fusion at
large-$p_T$, while the gluon fragmentation is negligibly small
over the whole range of $p_T$. An experimental study of $p_T$
distributions of $J/\psi$ at HERA will provide a direct
probe of the charm quark fragmentation functions.
\vspace{98pt}
\noindent
\begin{flushleft}
November 1995\\
\end{flushleft}

\vskip 10pt

\setcounter{footnote}{0}
\renewcommand{\thefootnote}{\arabic{footnote}}

\vfill
\clearpage
\setcounter{page}{1}
\pagestyle{plain}
The anomalously large cross-section for $J/\psi$ production at large
transverse momentum, $p_T$, measured \cite{cdf} recently in the CDF
experiment at the Tevatron has led to a revision of
earlier ideas based on the lowest order QCD process of parton
fusion. In this approach \cite{berjon, br}, the dominant
contribution to quarkonium production was expected to come
from quark-antiquark or gluon-gluon fusion, leading to the
formation of a heavy-quark pair in a colour-singlet state
with the correct spin, parity and charge-conjugation assignments
projected out. Several recent works \cite{bryu, bryu2, chen, yuan}
have drawn attention to additional contributions to quarkonium
production coming from the fragmentation of gluons and heavy
quark jets. Even though the fragmentation contributions are of
higher order in $\alpha_s$ compared to fusion, they are enhanced
by powers of $p_T^2/m^2$, where $m$ is the heavy quark mass.
Consequently, they can overtake the fusion contribution at $p_T \gg m$.
Indeed, the CDF $J/\psi$ production data \cite{cdf} has been
successfully explained by several authors \cite{jpsi} by taking
into account both the fusion and fragmentation contributions.
The gluon fragmentation contribution is found to dominate over
fusion at large $p_T$ ($p_T > 5$~GeV), while the charm quark
fragmentation contribution is much too small. As we shall see
in this letter, the photoproduction of $J/\psi$ at HERA presents
a complementary process~-- i.e. the charm quark fragmentation is
expected to overtake fusion at large-$p_T$, while the gluon
fragmentation remains small. Thus a measurement of the
large-$p_T$ $J/\psi$ production cross-section at HERA will
provide a valuable probe for the charm quark fragmentation
contribution.

\vskip10pt
A brief discussion of the colour-singlet model, used in the
computation of both fusion and fragmentation contributions,
is in order. Strictly speaking, the colour-singlet model is
a non-relativistic model where the relative velocity between
the heavy quarks in the bound state is ignored. However, in general,
the relative velocity, $v$, in quarkonium systems is not negligible
and O$(v)$ corrections need to be taken into account. Starting from a
non-relativistic QCD Lagrangian, a systematic analysis using the
factorisation method has been recently carried
out by Bodwin, Braaten and Lepage \cite{bbl}. In this formulation,
the quarkonium wave-function admits of a systematic expansion in
powers of $v$ in terms of Fock-space components~: for example, the
wave-functions for the $P$-state charmonia have the conventional
colour-singlet $P$-state component at leading order, but there exist
additional contributions at non-leading order in $v$, which involve octet
$S$-state components; i.e.
\begin{equation}
  \vert \chi_J \rangle = O(1) \vert Q\bar Q \lbrack {}^3P_J^{(1)} \rbrack
      \rangle + O(v) \vert Q\bar Q \lbrack {}^3S_1^{(8)} \rbrack g
      \rangle + \ldots
      \label{e1}
\end{equation}
In spite of the fact that the octet component in the wave-function is
suppressed by a factor of $v$, it is important for the decays of
$P$-states \cite{bbl2} for the following reasons~: 1) The $P$-state
wave-function is already suppressed by a factor of $v$ owing to the
angular-momentum barrier; but the corresponding colour-octet component
is an $S$-state which is unhindered by this barrier. The colour-octet
component can, therefore, easily compete with the colour-singlet. 2) The
second reason is even more compelling~: the perturbative analyses of
$P$-wave decays of quarkonia \cite{bgr} reveal a logarithmic infrared
singularity. But the octet component allows the infrared singularity
to be absorbed via a wave-function renormalisation, without having
to introduce an arbitrary infrared cut-off. So a consistent perturbative
treatment of $P$-state decays necessarily involves the octet component.
The price to pay for this is that two independent matrix-elements, viz.,
the singlet and the octet matrix elements are needed,
unlike the case of the colour-singlet model where the entire long-distance
information could be factorised into a $single$ non-perturbative
matrix-element. As in the case of the $P$-state decay widths, the
$P$-state fragmentation functions also involve the octet component
\cite{bryu2}. The octet component appears in the computation of the
fusion contribution as well, but is negligible in the large-$p_T$
region of our interest.

\vskip10pt
For $S$-state resonances like the $J/\psi$ and the
$\psi^{\prime}$, the octet contribution is suppressed by powers of
$v$. Further, the $S$-wave amplitude is not infrared divergent
and can, therefore, be described in terms of a single colour-singlet
matrix-element. But recently, the CDF collaboration has measured \cite{cdf2}
the ratio of $J/\psi$'s coming from $\chi$ decays to those produced
directly and it turns out that the direct $S$-state production is much
larger than the theoretical estimate. It has been suggested \cite{cgmp}
that a colour octet component in the $S$-wave production coming from
gluon fragmentation as originally proposed in Ref.~\cite{brfl}, can
explain this $J/\psi$ anomaly.
This corresponds to a virtual gluon fragmenting into an octet ${}^3S_1$
state which then makes a double E1 transition into a singlet ${}^3S_1$
state. While this process is suppressed by a factor of $v^4$ as
compared to the colour-singlet process, it is enhanced by a factor
of $\alpha_s^2$. One can fix the value of the colour-octet matrix-element
by normalising to the data on direct $J/\psi$ production cross-section
from the
CDF experiment. The colour-octet contribution to $S$-state production
has also been invoked \cite{brfl} to explain the large $\psi^{\prime}$
cross-section measured by CDF \cite{cdf}, but there can be a large
contribution to this cross-section coming from the
decays of radially excited $P$-states \cite{psip}.
Independent tests of the $S$-state colour octet enhancement are
important and there have been recent suggestions \cite{sugg} as to how
one can use $e^+e^-$ collisions to probe the octet contribution. Thus,
the possibility of a large colour-octet contribution to the $S$-state
fragmentation function remains open, though theoretically less
compelling than for the $P$-state. We shall see below that the
inclusive photoproduction of $J/\psi$ is insensitive to the
former, but it is sensitive to the latter.

\vskip10pt
In this letter, we study inclusive $J/\psi$ production in $ep$
collisions at HERA. The fusion contribution
to the photoproduction of $J/\psi$ in the
colour-singlet model \cite{berjon} comes from photon-gluon
fusion. Recently, the next-to-leading order corrections
to this process have been computed within the colour-singlet model
\cite{kramer1} and compared \cite{kramer2} with the results on
integrated cross-sections from HERA; and it has been found that
the integrated cross-sections at next-to-leading order
are in reasonable agreement with the data.
The integrated cross-sections are, however, insensitive to the
fragmentation contributions, because the latter dominate only at large
$p_T$. To get a handle on the fragmentation contributions to
$J/\psi$ production at HERA it is important to study the $p_T$
distributions, rather than integrated cross-sections.

\vskip10pt
The fusion contribution to the
photoproduction of $J/\psi$ in the colour-singlet model takes place
through the following subprocess:
\begin{equation}
\gamma + g \rightarrow c \bar c \lbrack {}^3S_1^{(1)} \rbrack + g,
      \label{e2}
\end{equation}
where the $J/\psi$ is taken to be the colour-singlet ${}^3S_1$ $c \bar c$
state. The $p_T$ differential cross-section for the photoproduction
of $J/\psi$ in the colour-singlet model is given as
\begin{equation}
{d\sigma \over dp_T} = \int dz {128 \pi^2 \alpha_s^2 \alpha p_T xG(x)
     z(1-z) M e_c^2 R_0^2 \over 27 \lbrack M^2(1-z)+p_T^2\rbrack^2}
       \cdot f(z,p_T^2),
      \label{e3}
\end{equation}
where
\begin{eqnarray}
f(z,p_T^2) & = & {1 \over (M^2+p_T^2)^2}+{(1-z)^4 \over \lbrack
      p_T^2+M^2(1-z)^2 \rbrack^2} \nonumber  \\
   && + {z^4p_T^4 \over
   (M^2+P_T^2)^2 \lbrack p_T^2+M^2(1-z)^2 \rbrack^2}.
      \label{e4}
\end{eqnarray}
In the above equation, the variable $z$ is the inelasticity
variable, defined as
\begin{equation}
z={p_{\psi} \cdot p_p \over p_{\gamma} \cdot p_p},
      \label{e5}
\end{equation}
and $x$ is related to $p_T$ and $z$~,
\begin{equation}
x = {1 \over s} \biggl \lbrack {M^2 \over z} +
{p_T^2 \over z(1-z)} \biggr\rbrack ,
      \label{e6}
\end{equation}
where $s=4E_p\nu$ is the photon-proton c.m. energy. As usual,
$M$ and $R_0$ denote the $J/\psi$ mass and wave function at
the origin.

\vskip10pt
The fragmentation contribution is computed by factorising the
cross-section for the process $\gamma p \rightarrow (J/\psi,\chi_i) X$
into a part containing the hard-scattering cross-section for producing a
gluon or a charm quark and a part which specifies the fragmentation of
the gluon or the charm quark into the required charmonium state, i.e.
\begin{equation}
d\sigma (\gamma p \rightarrow (J/\psi,\chi_i) X)
 = \sum_c \int_0^1 d\omega \hskip4pt
d\sigma (\gamma p \rightarrow c X) D_{c \rightarrow (J/\psi,\chi_i)}
(\omega,\mu ) ,
\label{e7}
\end{equation}
where $c$ is the fragmenting parton (either a gluon or a charm quark).
$D(\omega,\mu)$ is the fragmentation function and $\omega$ is
the fraction of the momentum of the parent parton carried by the
charmonium state\footnote{We use the notation $\omega$ instead
of the more usual $z$ to avoid confusion with the inelasticity
parameter, defined in Eq.~\ref{e5}.}. The fragmentation function is
computed perturbatively at an initial scale $\mu_0$
which is of the order of $m_c$. If the scale $\mu$ is chosen to be
of the order of $p_T$, then large logarithms in $\mu/m_c$ appear
which have then to be resummed using the usual Altarelli-Parisi equation:
\begin{equation}
\mu {\partial \over \partial\mu} D_{i\rightarrow (J/\psi,\chi_i)}
(\omega) = \sum_j\int_{\omega}^1{dy \over y} P_{ij}({\omega \over y},\mu)
D_{j\rightarrow (J/\psi,\chi_i)}(y) ,
\label{e8}
\end{equation}
where the $P_{ij}$ are the splitting functions of a parton $j$
into a parton $i$. We consider the fragmentation of gluons and
charm quarks alone since the light quark contributions are expected
to be very small. The gluons are produced via the Compton process:
\begin{equation}
\gamma + q \rightarrow q + g,
      \label{e9}
\end{equation}
whereas the charm quarks are produced via the Bethe-Heitler process:
\begin{equation}
\gamma + g \rightarrow c + \bar c.
      \label{e10}
\end{equation}
Using these cross-sections, we compute the fragmentation contribution
to $d\sigma/dp_T$ which is given by a formula similar to Eq.~\ref{e3},
but with an extra integration over $\omega$, or equivalently over $x$,
because of the relation
\begin{equation}
\omega={M^2+p_T^2 \over xsz} + z.
    \label{e11}
\end{equation}
For the fragmentation functions at the initial scale $\mu=\mu_0$, we
use the results of Refs.~\cite{bryu} and \cite{bryu2} for the
gluon fragmentation functions
into $J/\psi$ or $\chi$ states, and Refs.~\cite{chen} and
\cite{yuan} for the corresponding fragmentation functions
of the charm quark. These
fragmentation functions include
the colour-octet component in the $P$-state, but do not
include any colour-octet contribution in the $S$-state. For the
case of gluon fragmentation, we have separately studied the effect
of the $S$-state colour-octet component by modifying the fragmentation
functions as in Ref.~\cite{brfl}. For the charm fragmentation,
the $S$-state colour-octet contributions are sub-dominant and we have
neglected these contributions.
In principle, at HERA energies we can also expect
contributions from $B$-decays but these turn out to be dominant
at values of $z \le 0.1$ \cite{mns}, and can, therefore, be
safely neglected in our analysis.

\vskip10pt
We have computed the cross-sections for two representative values
of the photon energy, $\nu$, using the MRSD-${}^\prime$
structure functions \cite{mrs} and we have used $Q=p_T/2$
as the choice of scale. In principle, one can integrate over
the photon energy spectrum; but for the purposes of studying the
relative magnitudes of the fusion and fragmentation contributions to
the cross-sections, it is enough and indeed more transparent to
present the results for fixed values of $\nu$. In Fig.~1, we present
the results for $d\sigma / dp_T$ as a function of $p_T$, for $\nu=40$
and 100~GeV. For the inelasticity parameter, we use the cuts $0.1 \le
z \le 0.9$, as used in the ZEUS experiment at HERA \cite{zeus}.
We find that the
fusion contribution, shown by the solid line in Fig.~1, is dominant
at low $p_T$, but the charm quark fragmentation contribution (shown
by the dashed-dotted line) becomes important for values of $p_T$
greater than about 10~GeV. The gluon fragmentation contribution
(shown by the dashed line in the figure) is smaller by over an
order of magnitude throughout the range of $p_T$ considered. Also
shown as the dotted line in the figure is the gluon fragmentation
contribution including the octet contribution for the $S$-state, where
the numerical value of the octet $S$-wave matrix element has been
determined \cite{cgmp,brfl} from the CDF data \cite{cdf2}
as mentioned above. The major uncertainty in
the prediction is due to the limited information we have on
the colour-octet matrix elements; the normalisation
of the fragmentation contribution can change by a factor of 2-3,
due to this uncertainty \cite{jpsi}. Moreover, next-to-leading order
corrections
will also change the absolute normalisation of our predictions~--
for the fusion prediction this is expected to give an
enhancement (K-factor) upto a factor of 2
\cite{kramer2} and similar K-factors are also expected
in the case of fragmentation contributions. However, our choice
of $p_T/2$ as the scale (instead of $p_T$) is expected to account
for the $K$-factor enhancement, at least in part.

\vskip10pt
The charm fragmentation subprocess (Eq.~\ref{e10})
is gluon-initiated while the gluon fragmentation
subprocess (Eq.~\ref{e9}) is quark-initiated. This explains why the
charm fragmentation process dominates. It is important to note
that gluon fragmentation turns out to be the most important source
of $J/\psi$ production at the Tevatron, while the complementary
information on the fragmentation of charm quarks can be studied at HERA.
An experimental study of $p_T$ distributions at HERA will provide
us with the first direct measurement of the charm quark fragmentation
functions.

\vskip10pt
Since the $J/\psi$'s produced in the fragmentation process are softer
in energy than those produced $via$ fusion, it turns out that the average
value of $z$ for the former are smaller than the latter. To enhance the
fragmentation contribution, it is efficient to use a stronger upper
cut on $z$. In Fig.~2, we have shown the cross-sections for $J/\psi$'s
produced via fusion and from charm quark fragmentation with
$z<0.5$. We find that this cut helps to cut down the
fusion contribution to $J/\psi$
without significantly affecting the charm fragmentation
contribution, thereby providing a better signal for the fragmentation
process.

\vskip10pt
To summarise, we have studied the $p_T$ distribution of $J/\psi$
cross-sections at HERA coming from the fusion and fragmentation
processes. We find that the large-$p_T$ end is dominated
by contributions from charm quark fragmentation. The information that
can be obtained from HERA is thus complementary to that obtained from
the large-$p_T$
$J/\psi$ production at the Tevatron, which is dominated by the gluon
fragmentation contribution. By applying judicious cuts on the
inelasticity parameter $z$ it is possible to enhance the charm
fragmentation contribution relative to the fusion contribution.

\newpage
\begin{figure}[htb]
\vskip 9.5in\relax\noindent\hskip -1in\relax{\includegraphics{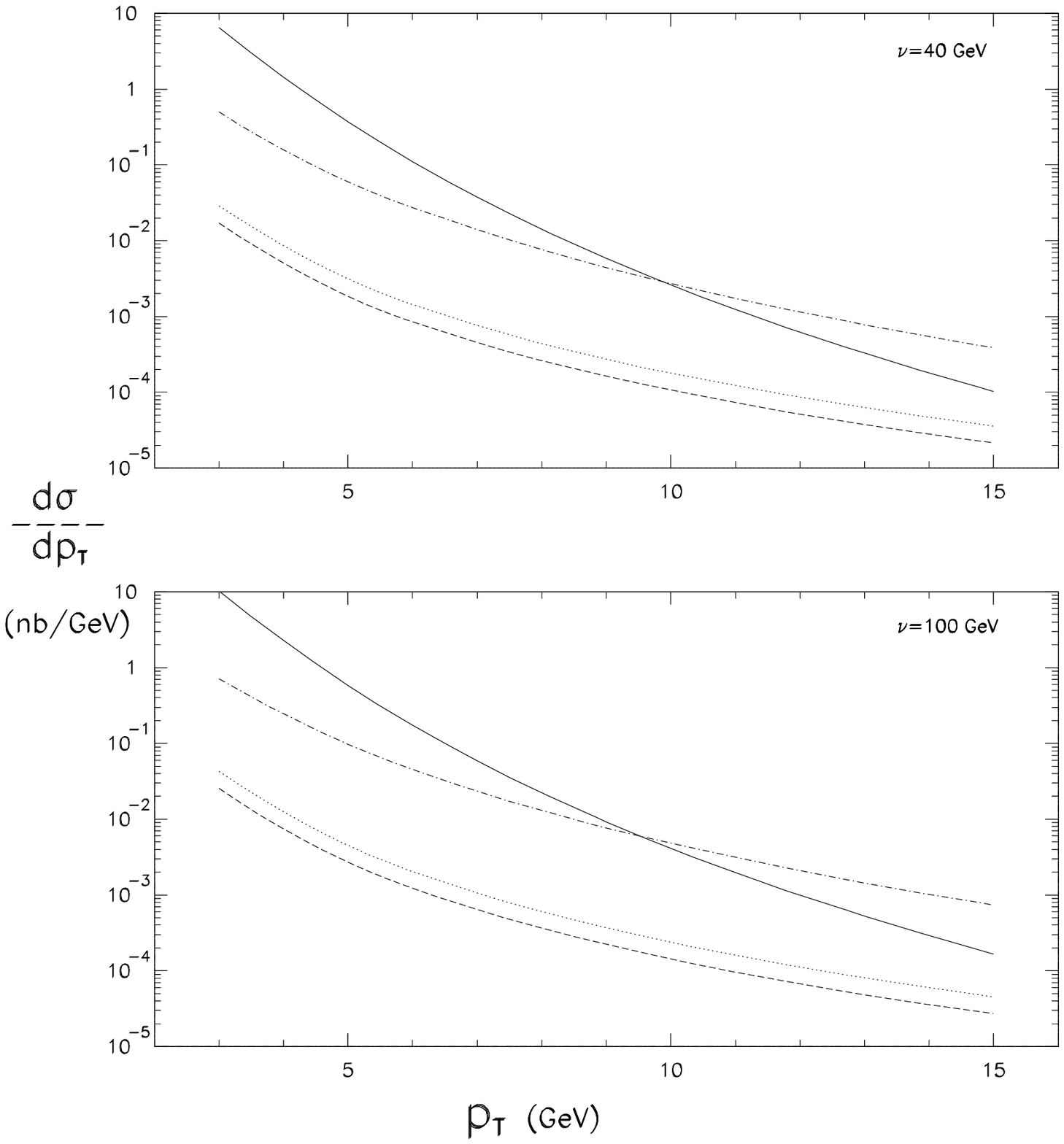}}

\vspace{-35ex}
\caption{$d\sigma/dp_T$ (in nb/GeV) for inclusive $J/\psi$ production
at HERA for photon energy $\nu=40$~GeV
(upper figure) and $\nu=100$~GeV (lower figure). The solid line
represents the fusion contribution
and the dashed-dotted line the charm quark fragmentation contribution.
The dotted and dashed lines
represent the gluon fragmentation contributions with and without a
colour-octet component for the $S$-state.
The cut on the inelasticity parameter, $z$, is $0.1 < z <0.9$.}
\end{figure}
\begin{figure}[htb]
\vskip 8in\relax\noindent\hskip -1in\relax{\includegraphics{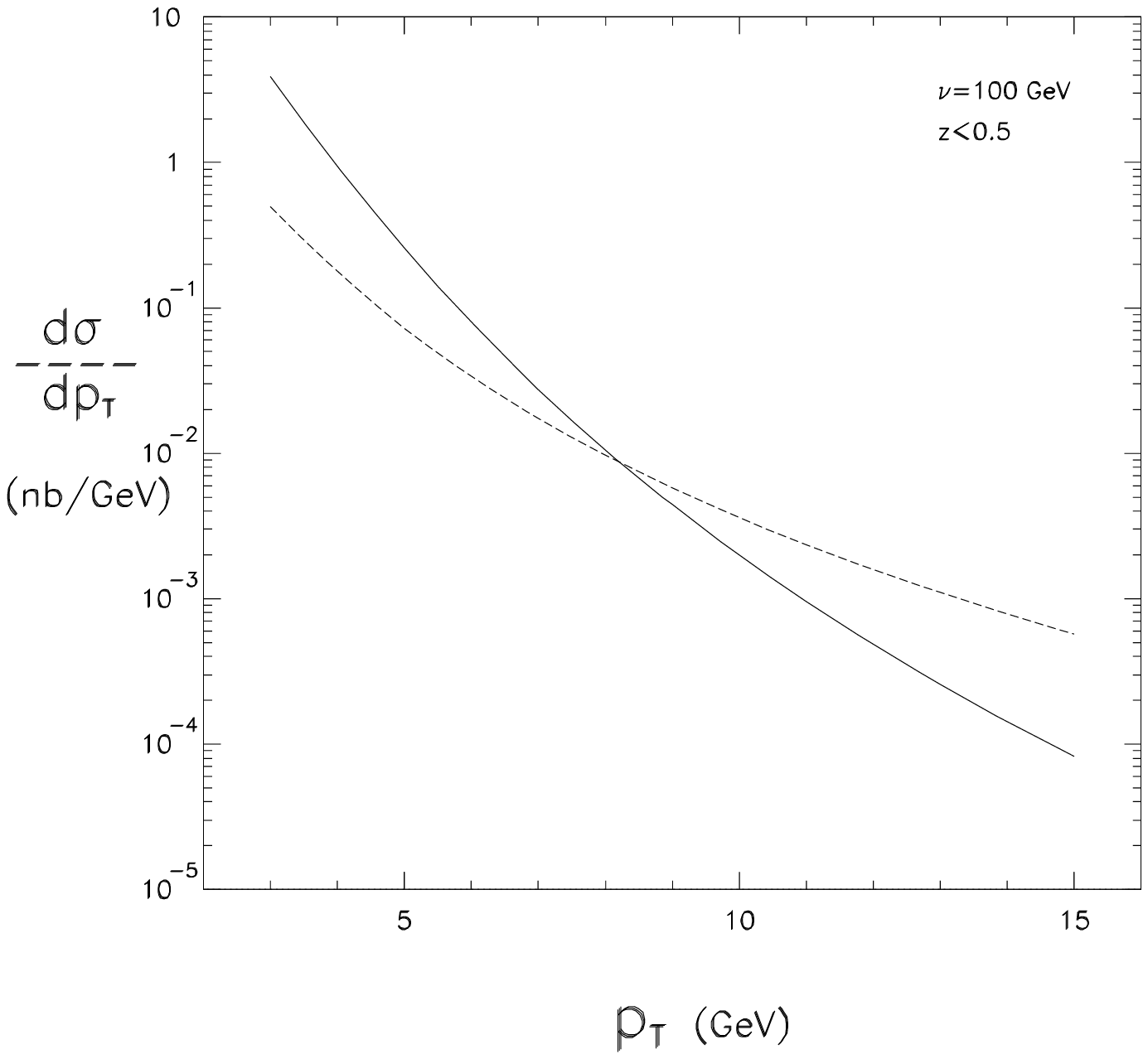}}

\vspace{-20ex}
\caption{$d\sigma/dp_T$ (in nb/GeV) for photon energy $\nu=100$~GeV.
The solid line represents the fusion contribution, and the
dashed line the charm quark
fragmentation contribution, using a cut $z < 0.5$.}
\end{figure}

\clearpage
\begin{thebibliography}{999}

\bibitem{cdf} F.~Abe et al., {\it Phys. Rev. Lett.} {\bf 69}
(1992) 3704; {\it Phys. Rev. Lett.} {\bf 71} (1993) 2537;
K.~Byrum, CDF Collaboration, Proceedings of the 27th International
Conference on High Energy Physics, Glasgow, (1994), eds. P.J.~Bussey
and I.G.~Knowles (Inst. of Physics Publ.) p.989.

\bibitem{berjon} E.L.~Berger and D.~Jones, {\it Phys. Rev.} {\bf D 23}
(1981) 1521.

\bibitem{br} R.~Baier and R.~R\" uckl, {\it Z. Phys.} {\bf C 19}
(1983) 251.

\bibitem{bryu} E.~Braaten and T.C.~Yuan, {\it Phys. Rev. Lett.}
{\bf 71} (1993) 1673.

\bibitem{bryu2} E.~Braaten and T.C.~Yuan,
{\it Phys. Rev.} {\bf D 50} (1994) 3176.

\bibitem{chen} Y.Q.~Chen, {\it Phys. Rev.} {\bf D 48} (1993) 5181.

\bibitem{yuan} T.C.~Yuan, {\it Phys. Rev.} {\bf D 50} (1994) 5664.

\bibitem{jpsi} E.~Braaten, M.A.~Doncheski, S.~Fleming and M.~Mangano,
{\it Phys. Lett.} {\bf B 333} (1994) 548; D.P.~Roy and K.~Sridhar,
{\it Phys. Lett.} {\bf B 339} (1994) 141; M.~Cacciari and M.~Greco,
{\it Phys. Rev. Lett.} {\bf 73} (1994) 1586.

\bibitem{bbl} G.T.~Bodwin, E.~Braaten and G.P.~Lepage, {\it Phys. Rev.}
{\bf D 51} (1995) 1125.

\bibitem{bbl2} G.T.~Bodwin, E.~Braaten and G.P.~Lepage, {\it Phys. Rev.}
{\bf D 46} (1992) R1914.

\bibitem{bgr} R.~Barbieri, R.~Gatto and E.~Remiddi, {\it Phys. Lett.}
{\bf B 61} (1976) 465.

\bibitem{cdf2}
G.~Bauer, CDF Collaboration, presented at the "Xth Topical Workshop
on $p \bar p$ collisions", Fermilab (1995).

\bibitem{cgmp} M.~Cacciari, M.~Greco, M.~Mangano and A.~Petrelli,
CERN Preprint CERN-TH/95-129; P.~Cho and A.K.~Leibovich,
Caltech Preprint CALT-68-1988.

\bibitem{brfl} E.~Braaten and S.~Fleming, {\it Phys. Rev. Lett.}
{\bf 74} (1995) 3327.

\bibitem{psip} F.E.~Close, {\it Phys. Lett.} {\bf B 342} (1995) 369;
D.P.~Roy and K.~Sridhar, {\it Phys. Lett.} {\bf B 345} (1995) 537.

\bibitem{sugg} E.~Braaten and S.~Fleming,  Northwestern University
Preprint NUHEP-TH-95-9; K.~Cheung, W.~Keung and T.C.~Yuan, Fermilab
Preprint FERMILAB-PUB-95/300-T; P.~Cho, Caltech Preprint CALT-68-2020.

\bibitem{kramer1} M.~Kr\" amer, J.~Zunft, J.~Steegborn and P.M.~Zerwas,
{\it Phys. Lett.} {\bf B 348} (1995) 657.

\bibitem{kramer2} M.~Kr\" amer, DESY Preprint DESY 95-155.

\bibitem{mns} A.D.~Martin, C.-K.~Ng and W.J.~Stirling, {\it Phys. Lett.}
{\bf B 191} (1987) 200.

\bibitem{mrs} A.D.~Martin, R.G.~Roberts and W.J.~Stirling,
{\it Phys. Lett.} {\bf B 306} (1993) 145;
{\it Phys. Lett.} {\bf B 309} (1993) 492.

\bibitem{zeus} M.~Derrick et al., contribution to the International
Europhysics Conference on High Energy Physics, Brussels, 1995.

\end{thebibliography}
\end{document}